\documentclass{article}
\usepackage{amssymb,amsmath,bm,hyperref}
\newtheorem{theorem}{Theorem}
\newtheorem{lemma}{Lemma}
\newtheorem{remark}{Remark}

\begin{document}

\title{Massless Dirac equation as a special case of Cosserat elasticity}

\author{
Olga Chervova\thanks{
Department of Mathematics,
University College London,
Gower Street,
London WC1E~6BT, UK;
olgac@math.ucl.ac.uk.
} \ and
Dmitri Vassiliev\thanks{
Department of Mathematics and Institute of Origins,
University College London,
Gower Street,
London WC1E~6BT, UK;
D.Vassiliev@ucl.ac.uk.
}
}

\maketitle

\vspace{-25pt}

\begin{abstract}
We suggest an alternative mathematical model for the massless
neutrino. Consider an elastic continuum in 3-dimensional Euclidean
space and assume that points of this continuum can experience no
displacements, only rotations. This framework is a special case of
the so-called Cosserat theory of elasticity. Rotations of points of
the continuum are described by attaching to each point an
orthonormal basis which gives a field of orthonormal bases called
the coframe. As the dynamical variables (unknowns) of our theory we
choose a coframe and a density. We write down a potential energy
which is conformally invariant and then incorporate time in the
standard Newtonian way, by subtracting kinetic energy. Finally, we
rewrite the resulting nonlinear variational problem in terms of an
unknown spinor field. We look for quasi-stationary
solutions, i.e. solutions that harmonically oscillate in time. We
prove that in the quasi-stationary setting our model is equivalent
to a pair of massless Dirac equations. The crucial element of the
proof is the observation that our Lagrangian admits a factorisation.
\end{abstract}

\section{Introduction}

The massless Dirac equation is a system of two homogeneous linear
complex partial differential equations for two complex unknowns. The
unknowns (components of a spinor) are functions of time and the
three spatial coordinates. This equation is the accepted
mathematical model for the massless neutrino.

The geometric interpretation of the massless Dirac equation is rather
complicated. It relies on the use of notions such as
\begin{itemize}
\item
spinor,
\item
Pauli matrices,
\item
covariant derivative (note that formula
(\ref{covariant derivative of spinor field})
for the covariant
derivative of a spinor field is quite tricky).
\end{itemize}
There is also a logical problem with the massless Dirac equation in
that it predicts the existence of four essentially different types
of plane wave solutions which are called left-handed neutrino,
right-handed neutrino, left-handed antineutrino and right-handed
antineutrino. Only left-handed neutrinos and right-handed
antineutrinos are observed experimentally.

The purpose of this paper is to formulate an alternative
mathematical model for the massless neutrino, a model which is
geometrically much simpler. The advantage of our approach is that it
does not require the use of spinors, Pauli matrices or covariant
differentiation. The only geometric concepts we use are those of a
\begin{itemize}
\item
metric,
\item
differential form,
\item
wedge product,
\item
exterior derivative.
\end{itemize}
Our model also overcomes the logical problem mentioned in the previous
paragraph in that it predicts the existence of only two essentially
different types of plane wave solutions. These correspond to
clockwise and anticlockwise rotations of the coframe.

The paper has the following structure.
In Section \ref{Notation and conventions} we introduce our notation and
in Section \ref{The Dirac equation} we formulate the massless Dirac equation.
In Section \ref{Our model} we formulate our mathematical model.
In Section \ref{Relativistic}
we rewrite our Lagrangian in relativistic form; we do this to show
that our Lagrangian looks simpler in relativistic notation, though we
do not pursue the relativistic approach consistently in the current paper.
In Section \ref{Choosing a common language} we translate our model
into the language of spinors.
In Section \ref{Quasi-stationary case}
we prove Theorem \ref{main theorem} which is the main result of
the paper: this theorem
establishes that in the
quasi-stationary case (prescribed oscillation in time with frequency $\omega$)
our mathematical model is equivalent to a pair
of massless Dirac equations.
The crucial element of the proof of Theorem \ref{main theorem}
is the observation that our Lagrangian admits a
factorisation; this factorisation is the subject of Lemma \ref{lemma 2}.
Section \ref{Plane wave solutions} deals with plane wave solutions.
The concluding discussion is contained in Section \ref{Discussion}.

\section{Notation and conventions}
\label{Notation and conventions}

We work on a 3-manifold $M$ equipped with prescribed
\textbf{negative} (i.e. with signature $---$) metric $g_{\alpha\beta}$.
We choose negative metric on the 3-manifold $M$
in order to facilitate the subsequent
extension (see Section \ref{Relativistic}) to a Lorentzian metric
\begin{equation}
\label{Lorentzian metric}
\mathbf{g}_{{\bm{\alpha}}{\bm{\beta}}}=
\begin{pmatrix}
1&0\\
0&g_{\alpha\beta}
\end{pmatrix}
\end{equation}
of signature $+---$
on the 4-manifold $\mathbb{R}\times M$.
We denote time by $x^0$ and
local coordinates on $M$ by $x^\alpha$, $\alpha=1,2,3$.

All constructions presented in the
paper are local so we do not make a priori assumptions on the
geometric structure of $\{M,g\}$.

Our notation follows \cite{MR2176749,vassilievPRD,jmp}. In particular,
in line with the traditions of particle physics, we use Greek
letters to denote tensor (holonomic) indices. The only difference
with references \cite{MR2176749,vassilievPRD,jmp} is that, by default, we assume
tensor indices to run through the values 1, 2, 3 rather than 0, 1,
2, 3. The index 0 will be treated separately.

Details of our spinor notation are given in Appendix A of
\cite{MR2176749}. We choose the zeroth Pauli matrix to be the identity matrix,
\begin{equation}
\label{Notation and conventions equation 1}
\sigma^0{}_{a\dot b}=
\sigma^{0a\dot b}=
\begin{pmatrix}
1&0\\
0&1\end{pmatrix}.
\end{equation}
The defining relations for the three remaining Pauli matrices are
\begin{equation}
\label{Notation and conventions equation 2}
\sigma^\alpha{}_{a\dot b}\sigma^{\beta c\dot b}
+\sigma^\beta{}_{a\dot b}\sigma^{\alpha c\dot b}
=2g^{\alpha\beta}\delta_a{}^c,
\qquad\alpha,\beta=1,2,3,
\end{equation}
\begin{equation}
\label{Notation and conventions equation 3}
\sigma^\alpha{}_{a\dot b}\sigma^{0a\dot b}
=0,
\qquad\alpha=1,2,3.
\end{equation}
We assume that all our Pauli matrices do not depend on time $x^0$.
Formulae
(\ref{Notation and conventions equation 1})--(\ref{Notation and conventions equation 3})
mean that we are effectively working on the 4-manifold
$\mathbb{R}\times M$ equipped with Lorentzian metric
(\ref{Lorentzian metric}).

By $\nabla$ we denote the covariant derivative
on the 3-manifold $M$
with respect to
the Levi-Civita connection.
It acts on a vector field and a spinor field as
$
\nabla_\alpha v^\beta:=\partial_\alpha v^\beta
+\Gamma^\beta{}_{\alpha\gamma}v^\gamma
$
and
\begin{equation}
\label{covariant derivative of spinor field}
\nabla_\alpha\xi^a:=
\partial_\alpha\xi^a
+\frac14\sigma_\beta{}^{a\dot c}
(\partial_\alpha\sigma^\beta{}_{b\dot c}
+\Gamma^\beta{}_{\alpha\gamma}\sigma^\gamma{}_{b\dot c})\xi^b
\end{equation}
respectively, where $\partial_\alpha:=\partial/\partial x^\alpha$ and
\[
\Gamma^\beta{}_{\alpha\gamma}=
\left\{{{\beta}\atop{\alpha\gamma}}\right\}:=
\frac12g^{\beta\delta}
(\partial_\alpha g_{\gamma\delta}
+\partial_\gamma g_{\alpha\delta}
-\partial_\delta g_{\alpha\gamma})
\]
are the Christoffel symbols.
We also denote
$\partial_0:=\partial/\partial x^0$.

We identify differential forms with covariant antisymmetric tensors.
Given a pair of real covariant antisymmetric tensors $P$ and $Q$ of
rank $r$ we define their dot product as
$
P\cdot Q:=\frac1{r!}P_{\alpha_1\ldots\alpha_r}Q_{\beta_1\ldots\beta_r}
g^{\alpha_1\beta_1}\ldots g^{\alpha_r\beta_r}
$.
We also define $\|P\|^2:=P\cdot P$.

\section{The Dirac equation}
\label{The Dirac equation}

The following system of two complex linear partial differential equations on
$\mathbb{R}\times M$ for two complex unknowns
is known as the \emph{massless Dirac equation}:
\begin{equation}
\label{Dirac equation}
i(\pm\sigma^0{}_{a\dot b}\partial_0
+\sigma^\alpha{}_{a\dot b}\nabla_\alpha)\xi^a=0.
\end{equation}
Here $\xi$ is a spinor field which plays the role
of dynamical variable (unknown quantity)
and is a function of time $x^0\in\mathbb{R}$
and local coordinates $(x^1,x^2,x^3)$ on the 3-manifold $M$.
Summation in (\ref{Dirac equation}) is carried out over $\alpha=1,2,3$.
The two choices of sign in (\ref{Dirac equation}) give two
versions of the massless Dirac equation which differ by time reversal.
Thus, we have a pair of massless Dirac equations.

The corresponding Lagrangian density is
\begin{multline}
\label{Dirac Lagrangian}
L_\mathrm{Dir}^\pm(\xi):=
\frac i2
\bigl[\pm
(\bar\xi^{\dot b}\sigma^0{}_{a\dot b}\partial_0\xi^a
-
\xi^a\sigma^0{}_{a\dot b}\partial_0\bar\xi^{\dot b})
\\
+
(\bar\xi^{\dot b}\sigma^\alpha{}_{a\dot b}\nabla_\alpha\xi^a
-
\xi^a\sigma^\alpha{}_{a\dot b}\nabla_\alpha\bar\xi^{\dot b})
\bigr]
\sqrt{|\det g|}\,.
\end{multline}

Note that the massless Dirac equation and Lagrangian are often called
\emph{Weyl} equation and Lagrangian respectively.

\section{Our model}
\label{Our model}

A \emph{coframe} $\vartheta$ is a triplet of real
covector fields
$\vartheta^j\in T^*M$, $j=1,2,3$,
satisfying the constraint
\begin{equation}
\label{constraint for coframe}
g=
-\vartheta^1\otimes\vartheta^1
-\vartheta^2\otimes\vartheta^2
-\vartheta^3\otimes\vartheta^3.
\end{equation}
For the sake of clarity we repeat formula~(\ref{constraint for coframe})
giving the tensor indices explicitly:
$g_{\alpha\beta}=
-\vartheta^1_\alpha\vartheta^1_\beta
-\vartheta^2_\alpha\vartheta^2_\beta
-\vartheta^3_\alpha\vartheta^3_\beta$.
We assume our coframe to be right-handed,
i.e. we assume that $\det\vartheta^j_\alpha>0.$

Formula (\ref{constraint for coframe}) means that
the coframe is a field of orthonormal bases.
Of course, at every point of the 3-manifold $M$ the choice of
coframe is not unique: there are 3 real degrees of freedom in
choosing the coframe and any pair of coframes is related by an
orthogonal transformation.

As dynamical variables in our model we choose
a coframe $\vartheta$ and a positive density $\rho$.
These live on the 3-manifold
$M$ and are functions of local coordinates $(x^1,x^2,x^3)$
as well as of time $x^0\in\mathbb{R}$.

At a physical level choosing the coframe as an unknown quantity
means that we view our 3-manifold $M$ as an elastic continuum and
allow every point of this continuum to rotate, assuming that
rotations of different points are totally independent. These
rotations are described mathematically by attaching to each point a
coframe (=~orthonormal basis). The approach in which the coframe
plays the role of a dynamical variable is known as
\emph{teleparallelism} (=~absolute parallelism =~fernparallelismus).
This is a subject promoted by A.~Einstein and \'E.~Cartan
\cite{letters,unzicker-2005-,sauer}.

The idea of rotating points may seem exotic, however it has long
been accepted in continuum mechanics within the so-called Cosserat
theory of elasticity \cite{Co}. The Cosserat theory of elasticity
has been in existence since 1909 and appears under various names in
modern applied mathematics literature such as \emph{oriented
medium}, \emph{asymmetric elasticity}, \emph{micropolar elasticity},
\emph{micromorphic elasticity}, \emph{moment elasticity} etc.
Cosserat elasticity is closely related to the theory of
ferromagnetic materials \cite{Ba} and the theory of liquid crystals
\cite{Lin,Ba2}. As to teleparallelism, it is, effectively, a special
case of Cosserat elasticity: here the assumption is that the elastic
continuum experiences no displacements, only rotations.
It is interesting to note that when Cartan
started developing what eventually became modern differential geometry
he acknowledged~\cite{cartan1}
that he drew inspiration from the `beautiful' work
of the Cosserat brothers.

We define the 3-form
\begin{equation}
\label{Our model equation 1}
T^\mathrm{ax}:=-\frac13
(\vartheta^1\wedge d\vartheta^1
+\vartheta^2\wedge d\vartheta^2
+\vartheta^3\wedge d\vartheta^3)
\end{equation}
where $\,d\,$ denotes the exterior derivative. This 3-form is called
\emph{axial torsion of the teleparallel connection}. An explanation
of the geometric meaning of the latter phrase as well as a detailed
exposition of the application of torsion in field theory and the
history of the subject can be found in \cite{cartantorsionreview}.
For our purposes the 3-form (\ref{Our model equation 1}) is simply a measure
of deformations generated by rotations of points of the 3-manifold $M$.

Note that the 3-form (\ref{Our model equation 1}) has the remarkable
property of conformal covariance: if we rescale our coframe as
\begin{equation}
\label{rescaling coframe}
\vartheta^j\mapsto e^h\vartheta^j
\end{equation}
where $h:M\to\mathbb{R}$ is an arbitrary scalar function,
then our metric is scaled as
\begin{equation}
\label{rescaling metric}
g_{\alpha\beta}\mapsto e^{2h}g_{\alpha\beta}
\end{equation}
and our 3-form is scaled as
\begin{equation}
\label{rescaling axial torsion}
T^\mathrm{ax}\mapsto e^{2h}T^\mathrm{ax}
\end{equation}
without the derivatives of $h$ appearing.

We take the potential energy of our continuum to be
\begin{equation}
\label{Our model equation 2}
P(x^0):=-\int_M\|T^\mathrm{ax}\|^2\rho\,dx^1dx^2dx^3.
\end{equation}
We put a minus sign in the RHS of (\ref{Our model equation 2}) because the
metric on our 3-manifold $M$ is negative, see beginning of Section
\ref{Notation and conventions}. As a result, our potential energy is
nonnegative, as it should be.

It is easy to see that the potential energy
(\ref{Our model equation 2})
is conformally
invariant: it does not change if we rescale our coframe
as (\ref{rescaling coframe}) and
our density as
\begin{equation}
\label{rescaling density}
\rho\mapsto e^{2h}\rho.
\end{equation}
This follows from formulae
(\ref{rescaling axial torsion}), (\ref{rescaling metric})
and
$\|T^\mathrm{ax}\|^2=\frac1{3!}
T^\mathrm{ax}_{\alpha\beta\gamma}T^\mathrm{ax}_{\kappa\lambda\mu}
g^{\alpha\kappa}g^{\beta\lambda}g^{\gamma\mu}$.
Thus, the guiding principle in our choice of potential energy
(\ref{Our model equation 2}) is conformal invariance.

We take the kinetic energy of our continuum to be
\begin{equation}
\label{Our model equation 3}
K(x^0):=\int_M\|\dot\vartheta\|^2\rho\,dx^1dx^2dx^3
\end{equation}
where $\dot\vartheta$ is the 2-form
\begin{equation}
\label{Our model equation 4}
\dot\vartheta:=\frac13
(\vartheta^1\wedge\partial_0\vartheta^1
+\vartheta^2\wedge\partial_0\vartheta^2
+\vartheta^3\wedge\partial_0\vartheta^3).
\end{equation}
Unlike (\ref{Our model equation 2}), we did not put a minus sign in the RHS of
(\ref{Our model equation 3}): this is because we are now squaring a
2-form rather than a 3-form (the number 2 is even whereas the number
3 is odd). As a result, our kinetic energy is nonnegative, as it
should be.

It is easy to see that the 2-form (\ref{Our model equation 4}) is,
up to a constant factor, the Hodge dual of angular velocity, so
(\ref{Our model equation 3}) is the standard expression for the
kinetic energy of a homogeneous isotropic Cosserat continuum. One
should think of a collection of identical infinitesimal rotating solid
balls distributed with density $\rho$. We think in terms of balls
rather than ellipsoids because of the isotropy, i.e. we do not have
preferred axes of rotation.

We now combine the potential energy (\ref{Our model equation 2}) and
kinetic energy (\ref{Our model equation 3}) to form the action
(variational functional) of our dynamic problem:
\begin{equation}
\label{Our model equation 5}
S:=\int_{\mathbb{R}}(K(x^0)-P(x^0))\,dx^0=
\int_{{\mathbb{R}}\times M}L(\vartheta,\rho)\,dx^0dx^1dx^2dx^3
\end{equation}
where
\begin{equation}
\label{Our model equation 6}
L(\vartheta,\rho):=(\|\dot\vartheta\|^2+\|T^\mathrm{ax}\|^2)\rho
\end{equation}
is our Lagrangian density. Note that this Lagrangian density
is conformally invariant in the Lorentzian sense.
The latter means that we rescale time
simultaneously with the rescaling of the 3-dimensional coframe.

Our field equations (Euler--Lagrange equations) are obtained by
varying the action (\ref{Our model equation 5}) with respect to the
coframe $\vartheta$ and density $\rho$. Varying with respect to the
density $\rho$ is easy: this gives the field equation
$\|\dot\vartheta\|^2+\|T^\mathrm{ax}\|^2=0$ which is equivalent to
$L(\vartheta,\rho)=0$. Varying with respect to the coframe
$\vartheta$ is more difficult because we have to maintain the metric
constraint (\ref{constraint for coframe}); recall that the metric is
assumed to be prescribed (fixed).

We do not write down the field equations for the Lagrangian density
$L(\vartheta,\rho)$ explicitly. We note only that they are highly
nonlinear and do not appear to bear any resemblance to the linear Dirac
equation (\ref{Dirac equation}).

\section{Relativistic representation of our Lagrangian}
\label{Relativistic}

In this section we work on the 4-dimensional manifold
$\mathbb{R}\times M$ equipped with Lorentzian metric
(\ref{Lorentzian metric}). This manifold is an extension of the
original 3-manifold~$M$. We use \textbf{bold} type for extended quantities.

We extend our coframe as
\begin{equation}
\label{Relativistic equation 1}
{\bm{\vartheta}}{}^0_{\bm{\alpha}}=
\begin{pmatrix}1\\0_\alpha\end{pmatrix},
\end{equation}
\begin{equation}
\label{Relativistic equation 2}
{\bm{\vartheta}}{}^j_{\bm{\alpha}}=
\begin{pmatrix}0\\\vartheta^j_\alpha\end{pmatrix},
\qquad j=1,2,3,
\end{equation}
where the bold tensor index $\bm{\alpha}$ runs through the values 0,
1, 2, 3, whereas its non-bold counterpart $\alpha$ runs through the
values 1, 2, 3. In particular, the $0_\alpha$ in formula
(\ref{Relativistic equation 1}) stands for a column of three zeros.

The extended metric (\ref{Lorentzian metric}) is expressed via the extended
coframe (\ref{Relativistic equation 1}), (\ref{Relativistic equation 2}) as
\begin{equation}
\label{Relativistic equation 3}
\mathbf{g}=
{\bm{\vartheta}}{}^0\otimes{\bm{\vartheta}}{}^0
-{\bm{\vartheta}}{}^1\otimes{\bm{\vartheta}}{}^1
-{\bm{\vartheta}}{}^2\otimes{\bm{\vartheta}}{}^2
-{\bm{\vartheta}}{}^3\otimes{\bm{\vartheta}}{}^3
\end{equation}
(compare with (\ref{constraint for coframe})).
The extended axial torsion is
\begin{equation}
\label{Relativistic equation 4}
\mathbf{T}^\mathrm{ax}:=\frac13
(
\underset{=0}
{
\underbrace{
{\bm{\vartheta}}{}^0\wedge d{\bm{\vartheta}}{}^0
}
}
-{\bm{\vartheta}}{}^1\wedge d{\bm{\vartheta}}{}^1
-{\bm{\vartheta}}{}^2\wedge d{\bm{\vartheta}}{}^2
-{\bm{\vartheta}}{}^3\wedge d{\bm{\vartheta}}{}^3
)
\end{equation}
where $\,d\,$ denotes the exterior derivative on $\mathbb{R}\times M$.
Formula (\ref{Relativistic equation 4}) can be rewritten~as
\begin{equation}
\label{Relativistic equation 5}
\mathbf{T}^\mathrm{ax}
={\bm{\vartheta}}{}^0\wedge\dot\vartheta
+T^\mathrm{ax}
\end{equation}
with $\dot\vartheta$ and $T^\mathrm{ax}$ defined by
(\ref{Our model equation 4}) and (\ref{Our model equation 1}) respectively.
Squaring (\ref{Relativistic equation 5}) we get
$\|\mathbf{T}^\mathrm{ax}\|^2=\|\dot\vartheta\|^2+\|T^\mathrm{ax}\|^2$
which implies that our Lagrangian density (\ref{Our model equation 6})
can be rewritten as
\begin{equation}
\label{Relativistic equation 6}
L(\vartheta,\rho)=\|\mathbf{T}^\mathrm{ax}\|^2\rho.
\end{equation}

The point of the arguments presented in this section was to show that
if one accepts the relativistic point of view then our Lagrangian density
(\ref{Our model equation 6}) takes the especially simple form
(\ref{Relativistic equation 6}). A consistent pursuit of the relativistic
approach would require the variation of all four elements of the extended
coframe which we do not do in the current paper. Instead, we assume that
the zeroth element of the extended coframe is specified by formula
(\ref{Relativistic equation 1}).

\section{Choosing a common language}
\label{Choosing a common language}

In order to compare the two models described in Sections
\ref{The Dirac equation} and \ref{Our model}
we need to choose a common mathematical language.
We choose the language of spinors.
Namely, we express the coframe and density via a spinor field $\xi^a$
according to formulae
\begin{equation}
\label{common language equation 1}
s=\xi^a\sigma^0{}_{a\dot b}\bar\xi^{\dot b},
\end{equation}
\begin{equation}
\label{common language equation 2}
\rho=s\,\sqrt{|\det g|}\,,
\end{equation}
\begin{equation}
\label{common language equation 3}
(\vartheta^1+i\vartheta^2)_\alpha
=-s^{-1}\xi^a\sigma_{\alpha a\dot b}
\epsilon^{\dot b\dot c}\sigma^0{}_{d\dot c}\xi^d,
\end{equation}
\begin{equation}
\label{common language equation 4}
\vartheta^3_\alpha
=s^{-1}\xi^a\sigma_{\alpha a\dot b}\bar\xi^{\dot b}
\end{equation}
where
\begin{equation}
\label{common language equation 5}
\epsilon_{ab}=\epsilon_{\dot a\dot b}=
\epsilon^{ab}=\epsilon^{\dot a\dot b}=
\begin{pmatrix}
0&1\\
-1&0
\end{pmatrix}.
\end{equation}
Note that throughout this paper we assume that the density $\rho$
does not vanish. This is equivalent to the assumption that the spinor
field $\xi^a$ does not vanish.

Formulae (\ref{common language equation 1})--(\ref{common language equation 5})
are effectively a special case of those from Section 5 of \cite{rome},
the only difference being that now the two spinors in the bispinor are not
independent but related
as $\eta_{\dot b}=\sigma^0{}_{a\dot b}\xi^a$.
This is hardly surprising as in the current paper we work in a non-relativistic
3-dimensional setting so we do not really need bispinors.
We also do not really need to distinguish between dotted and undotted spinor
indices but we retained this distinction in order to facilitate comparison
with \cite{rome}.

Formulae (\ref{common language equation 1})--(\ref{common language equation 5})
establish a one-to-two correspondence between a coframe $\vartheta$
and a (positive) density $\rho$ on the one hand and a nonvanishing spinor field
$\xi^a$ on the other. The correspondence is one-to-two because
for given $\vartheta$ and $\rho$
the above formulae define $\xi^a$ uniquely up to choice of sign.
This is in agreement with the generally accepted view
(see, for example, Section 19 in \cite{LL4}
or Section 3.5 in \cite{MR1884336})
that the sign of a spinor does not have a physical meaning.

\begin{remark}
\label{rigid rotations of coframe}
Formulae
(\ref{common language equation 1})--(\ref{common language equation 5})
look somewhat
unnatural in that they are assign a special meaning to the element
$\vartheta^3$ of our coframe. This can be overcome by allowing rigid
rotations of the coframe, i.e. linear transformations
$\vartheta^j\mapsto O^j{}_k\vartheta^k$ where $O^j{}_k$ $j,k=1,2,3$,
is a constant orthogonal matrix with determinant $+1$. Note that such
transformations are totally unrelated to coordinate transformations.
It is well know that axial torsion is invariant under rigid rotations of
the coframe, hence our model described in Section
\ref{Our model} is invariant under rigid rotations of
the coframe. In particular, it is natural to view coframes which differ
by a rigid rotation as equivalent.
\end{remark}

\section{Quasi-stationary case}
\label{Quasi-stationary case}

For both models,
the traditional one (described in Section \ref{The Dirac equation})
and our model (described in Section \ref{Our model}),
we shall now seek solutions of the form
\begin{equation}
\label{Quasi-stationary case equation 1}
\xi^a(x^0,x^1,x^2,x^3)=
e^{-i\omega x^0}\zeta^a(x^1,x^2,x^3)
\end{equation}
where $\omega\ne0$ is a fixed real number.
We shall call solutions of the type
(\ref{Quasi-stationary case equation 1}) \emph{quasi-stationary}.
In seeking quasi-stationary solutions
what we are doing is separating out the time variable $x^0$,
as is done when reducing, say, the wave equation to the
Helmholtz equation.

Substituting (\ref{Quasi-stationary case equation 1}) into
(\ref{Dirac Lagrangian}) we get
\begin{equation}
\label{Quasi-stationary case equation 2}
L_\mathrm{Dir}^\pm(\zeta)=
\left[
\frac i2
(\bar\zeta^{\dot b}\sigma^\alpha{}_{a\dot b}\nabla_\alpha\zeta^a
-
\zeta^a\sigma^\alpha{}_{a\dot b}\nabla_\alpha\bar\zeta^{\dot b})
\pm\omega\zeta^a\sigma^0{}_{a\dot b}\bar\zeta^{\dot b}
\right]
\sqrt{|\det g|}\,.
\end{equation}

Substituting (\ref{Quasi-stationary case equation 1}) into
(\ref{common language equation 1})--(\ref{common language equation 5})
and the latter into (\ref{Our model equation 4}) we get
$\dot\vartheta=\frac43\omega\vartheta^1\wedge\vartheta^2$.
Hence, $\|\dot\vartheta\|^2=\frac{16}9\omega^2$ and formula
(\ref{Our model equation 6}) becomes
\begin{equation}
\label{Quasi-stationary case equation 3}
L(\zeta)=\left(\|T^\mathrm{ax}\|^2+\frac{16}9\omega^2\right)\rho
\end{equation}
where
\begin{equation}
\label{Quasi-stationary case equation 4}
\rho=\zeta^a\sigma^0{}_{a\dot b}\bar\zeta^{\dot b}\sqrt{|\det g|}\,.
\end{equation}
Note that our 3-dimensional metric is negative (see beginning of
Section \ref{Notation and conventions}),
so $\|T^\mathrm{ax}\|^2\le0$ and the Lagrangian density
(\ref{Quasi-stationary case equation 3}) may vanish. In fact, as we
shall see from the proof of Theorem \ref{main theorem} in the end of this section,
it has to vanish on solutions of our field equations.

In order to compare the Lagrangian densities
(\ref{Quasi-stationary case equation 2})
and
(\ref{Quasi-stationary case equation 3})
we need an explicit formula for $T^\mathrm{ax}$ in terms of the spinor
field. This formula is given in the following

\begin{lemma}
\label{lemma 1}
We have
\begin{equation}
\label{lemma 1 equation 1}
(*T^\mathrm{ax})\rho=
\frac{2i}3
(\bar\xi^{\dot b}\sigma^\alpha{}_{a\dot b}\nabla_\alpha\xi^a
-
\xi^a\sigma^\alpha{}_{a\dot b}\nabla_\alpha\bar\xi^{\dot b})
\sqrt{|\det g|}
\end{equation}
where
\begin{equation}
\label{lemma 1 equation 2}
*T^\mathrm{ax}:=
\frac1{3!}\,\sqrt{|\det g|}\,
(T^\mathrm{ax})^{\alpha\beta\gamma}\varepsilon_{\alpha\beta\gamma}
\end{equation}
is the Hodge dual of $T^\mathrm{ax}$, $\varepsilon_{123}:=+1$.
\end{lemma}

\textbf{Proof\ } Observe that time does not appear in the formula
(\ref{Our model equation 1}) for axial torsion
(summation is carried out over $\alpha=1,2,3$).
Hence the result we
are proving should hold for spinor fields $\xi$ with arbitrary
dependence on time and not only for quasi-stationary ones. As the
expression in the RHS of (\ref{lemma 1 equation 1}) is an invariant
we can also temporarily (for the duration of the proof of Lemma
\ref{lemma 1}) suspend our convention
(Section \ref{Notation and conventions})
that our Pauli matrices do not depend on time.

In order to simplify calculations we observe
that we have freedom in our choice of Pauli matrices. It is sufficient
to prove formula (\ref{lemma 1 equation 1}) for one particular
choice of Pauli matrices, hence it is natural to choose Pauli matrices in
a way that makes calculations as simple as possible.
Note that this trick is not new: it was, for example, extensively used by
A.~Dimakis and F.~M\"uller-Hoissen
\cite{muellerhoissen1,muellerhoissen2,muellerhoissen3}.

We choose Pauli matrices
\begin{equation}
\label{special formula for Pauli matrices}
\sigma_{\alpha a\dot b}=\vartheta^j_\alpha\,s_{ja\dot b}
=\vartheta^1_\alpha\,s_{1a\dot b}
+\vartheta^2_\alpha\,s_{2a\dot b}
+\vartheta^3_\alpha\,s_{3a\dot b}
\end{equation}
where
\begin{equation}
\label{Pauli matrices s}
s_{ja\dot b}=
\begin{pmatrix}
s_{1a\dot b}\\
s_{2a\dot b}\\
s_{3a\dot b}
\end{pmatrix}
:=
\begin{pmatrix}
\begin{pmatrix}
0&1\\
1&0
\end{pmatrix}
\\
\begin{pmatrix}
0&i\\
-i&0
\end{pmatrix}
\\
\begin{pmatrix}
1&0\\
0&-1\end{pmatrix}
\end{pmatrix}.
\end{equation}
Let us stress that in the statement of Lemma \ref{lemma 1} Pauli
matrices are not assumed to be related in any way to the coframe
$\vartheta$. We are just choosing the particular Pauli matrices
(\ref{special formula for Pauli matrices}), (\ref{Pauli matrices s})
to simplify calculations in our proof.

Substituting
(\ref{special formula for Pauli matrices}), (\ref{Pauli matrices s})
into
(\ref{common language equation 3}), (\ref{common language equation 4})
we see that the system
(\ref{common language equation 1})--(\ref{common language equation 5})
can be easily resolved for $\xi$:
solutions are spinors with
$\xi^2=0$ and $\xi^1$ which is nonzero and real.
Thus, we have
\begin{equation}
\label{special formula for spinor}
\xi^a=\pm
\begin{pmatrix}
e^h\\
0
\end{pmatrix}
\end{equation}
where $h:M\to\mathbb{R}$ is a scalar function.
Formula (\ref{special formula for spinor}) may seem strange: we are
proving Lemma \ref{lemma 1} for a general nonvanishing spinor field
$\xi$ but ended up with formula (\ref{special formula for spinor})
which is very specific. However, there is no contradiction here
because we chose Pauli matrices specially adapted to the coframe
and, hence, specially adapted to the corresponding spinor field.

Formulae (\ref{covariant derivative of spinor field})
and (\ref{special formula for spinor}) imply
\begin{multline*}
\frac i2
\bar\xi^{\dot d}\sigma^\alpha{}_{a\dot d}\nabla_\alpha\xi^a
=\frac i2
\bar\xi^{\dot d}(\sigma^\alpha{}_{a\dot d}\partial_\alpha h)\xi^a
+\frac i8\bar\xi^{\dot d}\sigma^\alpha{}_{a\dot d}
\sigma_\beta{}^{a\dot c}
(\partial_\alpha\sigma^\beta{}_{b\dot c}
+\Gamma^\beta{}_{\alpha\gamma}\sigma^\gamma{}_{b\dot c})\xi^b
\\
=\frac{ie^{2h}}8\sigma^\alpha{}_{a\dot 1}
\sigma_\beta{}^{a\dot c}
(\partial_\alpha\sigma^\beta{}_{1\dot c}
+\Gamma^\beta{}_{\alpha\gamma}\sigma^\gamma{}_{1\dot c})+\ldots
=\frac{is}8\sigma^\alpha{}_{a\dot 1}
\sigma_\beta{}^{a\dot c}
\nabla_\alpha\sigma^\beta{}_{1\dot c}+\ldots
\end{multline*}
where $s=\xi^a\sigma^0{}_{a\dot b}\bar\xi^{\dot b}=e^{2h}$
is the scalar (\ref{common language equation 1}) and
the ``dots'' denote purely imaginary terms.
We now write down the spinor summation indices explicitly:
\begin{multline*}
\frac i2
\bar\xi^{\dot d}\sigma^\alpha{}_{a\dot d}\nabla_\alpha\xi^a
=\frac{is}8
\bigl[\sigma^\alpha{}_{1\dot 1}
\sigma_\beta{}^{1\dot 1}
\nabla_\alpha\sigma^\beta{}_{1\dot 1}
+\sigma^\alpha{}_{1\dot 1}
\sigma_\beta{}^{1\dot 2}
\nabla_\alpha\sigma^\beta{}_{1\dot 2}
\\
+\sigma^\alpha{}_{2\dot 1}
\sigma_\beta{}^{2\dot 1}
\nabla_\alpha\sigma^\beta{}_{1\dot 1}
+\sigma^\alpha{}_{2\dot 1}
\sigma_\beta{}^{2\dot 2}
\nabla_\alpha\sigma^\beta{}_{1\dot 2}\bigr]+\ldots
\end{multline*}
Finally, substituting explicit formulae
(\ref{special formula for Pauli matrices}),
(\ref{Pauli matrices s})
for our Pauli matrices
and using the ``dots''
to absorb all purely imaginary terms we get
\begin{multline*}
\frac i2
\bar\xi^{\dot d}\sigma^\alpha{}_{a\dot d}\nabla_\alpha\xi^a
=\frac{is}8
\bigl[\vartheta^{3\alpha}
(-\vartheta^3_\beta)
\nabla_\alpha\vartheta^{3\beta}
-\vartheta^{3\alpha}
(\vartheta^1-i\vartheta^2)_\beta
\nabla_\alpha(\vartheta^1+i\vartheta^2)^\beta
\\
-(\vartheta^1-i\vartheta^2)^\alpha
(\vartheta^1+i\vartheta^2)_\beta
\nabla_\alpha\vartheta^{3\beta}
+(\vartheta^1-i\vartheta^2)^\alpha
\vartheta^3_\beta
\nabla_\alpha(\vartheta^1+i\vartheta^2)^\beta\bigr]+\ldots
\\
=\frac{is}8
\bigl[
-i\vartheta^{3\alpha}\vartheta^1_\beta\nabla_\alpha\vartheta^{2\beta}
+i\vartheta^{3\alpha}\vartheta^2_\beta\nabla_\alpha\vartheta^{1\beta}
-i\vartheta^{1\alpha}\vartheta^2_\beta\nabla_\alpha\vartheta^{3\beta}
\\
+i\vartheta^{2\alpha}\vartheta^1_\beta\nabla_\alpha\vartheta^{3\beta}
+i\vartheta^{1\alpha}\vartheta^3_\beta\nabla_\alpha\vartheta^{2\beta}
-i\vartheta^{2\alpha}\vartheta^3_\beta\nabla_\alpha\vartheta^{1\beta}\bigr]
+\ldots
\\
=\frac s8
\bigl[
\vartheta^{3\alpha}\vartheta^1_\beta\nabla_\alpha\vartheta^{2\beta}
-\vartheta^{3\alpha}\vartheta^2_\beta\nabla_\alpha\vartheta^{1\beta}
+\vartheta^{1\alpha}\vartheta^2_\beta\nabla_\alpha\vartheta^{3\beta}
\\
-\vartheta^{2\alpha}\vartheta^1_\beta\nabla_\alpha\vartheta^{3\beta}
-\vartheta^{1\alpha}\vartheta^3_\beta\nabla_\alpha\vartheta^{2\beta}
+\vartheta^{2\alpha}\vartheta^3_\beta\nabla_\alpha\vartheta^{1\beta}\bigr]
+\ldots
\\
=\frac s8
[(\vartheta^1\wedge\vartheta^2)\cdot d\vartheta^3
+(\vartheta^3\wedge\vartheta^1)\cdot d\vartheta^2
+(\vartheta^2\wedge\vartheta^3)\cdot d\vartheta^1]
+\ldots
\end{multline*}
Hence,
\begin{multline}
\label{lemma 1 equation 3}
\frac i2
(\bar\xi^{\dot b}\sigma^\alpha{}_{a\dot b}\nabla_\alpha\xi^a
-
\xi^a\sigma^\alpha{}_{a\dot b}\nabla_\alpha\bar\xi^{\dot b})
\\
=\frac s4
[(\vartheta^1\wedge\vartheta^2)\cdot d\vartheta^3
+(\vartheta^3\wedge\vartheta^1)\cdot d\vartheta^2
+(\vartheta^2\wedge\vartheta^3)\cdot d\vartheta^1].
\end{multline}

Axial torsion is defined by formula (\ref{Our model equation 1})
whereas
\[
\sqrt{|\det g|}\,\varepsilon_{\alpha\beta\gamma}
=(\vartheta^1\wedge\vartheta^2\wedge\vartheta^3)_{\alpha\beta\gamma}
\]
so formula (\ref{lemma 1 equation 2}) can be rewritten as
\begin{multline}
\label{lemma 1 equation 4}
*T^\mathrm{ax}
=T^\mathrm{ax}
\cdot(\vartheta^1\wedge\vartheta^2\wedge\vartheta^3)
=-\frac13
(\vartheta^1\wedge d\vartheta^1
+\vartheta^2\wedge d\vartheta^2
+\vartheta^3\wedge d\vartheta^3)
\cdot(\vartheta^1\wedge\vartheta^2\wedge\vartheta^3)
\\
=\frac13
[(\vartheta^1\wedge\vartheta^2)\cdot d\vartheta^3
+(\vartheta^3\wedge\vartheta^1)\cdot d\vartheta^2
+(\vartheta^2\wedge\vartheta^3)\cdot d\vartheta^1].
\end{multline}
Here we used the fact that because of the negativity of our metric
(see (\ref{constraint for coframe}))
we have $\|\vartheta^j\|^2=-1$, $j=1,2,3$.

Formulae (\ref{lemma 1 equation 4}),
(\ref{lemma 1 equation 3})
and (\ref{common language equation 2})
imply (\ref{lemma 1 equation 1}). $\square$

\

We are now in a position to establish the relationship between
the Lagrangian densities
(\ref{Quasi-stationary case equation 2})
and
(\ref{Quasi-stationary case equation 3}).

\begin{lemma}
\label{lemma 2}
In the quasi-stationary case (\ref{Quasi-stationary case equation 1})
our Lagrangian density (\ref{Quasi-stationary case equation 3}) factorises as
\begin{equation}
\label{lemma 2 equation 1}
L(\zeta)=-\frac{32\omega}9
\frac{L_\mathrm{Dir}^+(\zeta)L_\mathrm{Dir}^-(\zeta)}
{L_\mathrm{Dir}^+(\zeta)-L_\mathrm{Dir}^-(\zeta)}\,.
\end{equation}
\end{lemma}

Let us emphasise once again that throughout this paper we assume
that the density $\rho$ does not vanish.
In view of formulae
(\ref{Quasi-stationary case equation 2}),
(\ref{Quasi-stationary case equation 4}),
in the quasi-stationary case
this assumption can be
equivalently rewritten as
\begin{equation}
\label{density is nonzero}
L_\mathrm{Dir}^+(\zeta)\ne L_\mathrm{Dir}^-(\zeta)
\end{equation}
so the denominator in (\ref{lemma 2 equation 1}) is nonzero.

\textbf{Proof of Lemma \ref{lemma 2}\ }
In the quasi-stationary case
(\ref{Quasi-stationary case equation 1})
formula
(\ref{lemma 1 equation 1}) takes the form
\[
(*T^\mathrm{ax})\rho=
\frac{2i}3
\bigl[
(\bar\xi^{\dot b}\sigma^\alpha{}_{a\dot b}\nabla_\alpha\xi^a
-
\xi^a\sigma^\alpha{}_{a\dot b}\nabla_\alpha\bar\xi^{\dot b})
\bigr]
\sqrt{|\det g|}
\]
because the factor $e^{-i\omega x^0}$ cancels out.
Hence (recall that our metric is negative)
\[
\|T^\mathrm{ax}\|^2=
-\frac{16}{9\rho^2}
\left[
\frac i2
(\bar\xi^{\dot b}\sigma^\alpha{}_{a\dot b}\nabla_\alpha\xi^a
-
\xi^a\sigma^\alpha{}_{a\dot b}\nabla_\alpha\bar\xi^{\dot b})
\sqrt{|\det g|}\,
\right]^2.
\]
Substituting this into (\ref{Quasi-stationary case equation 3})
we get
\begin{equation}
\label{lemma 2 equation 2}
L(\zeta)=-\frac{16}{9\rho}
\left(
\left[
\frac i2
(\bar\xi^{\dot b}\sigma^\alpha{}_{a\dot b}\nabla_\alpha\xi^a
-
\xi^a\sigma^\alpha{}_{a\dot b}\nabla_\alpha\bar\xi^{\dot b})
\sqrt{|\det g|}\,
\right]^2-\omega^2\rho^2
\right).
\end{equation}
Formulae (\ref{lemma 2 equation 2}),
(\ref{Quasi-stationary case equation 2})
and (\ref{Quasi-stationary case equation 4})
imply (\ref{lemma 2 equation 1}). $\square$

\

The following theorem is the main result of our paper.

\begin{theorem}
\label{main theorem}
In the quasi-stationary case (\ref{Quasi-stationary case equation 1})
a spinor field $\zeta$ is a solution of the field
equations for the Lagrangian density $L(\zeta)$ if and only
if this spinor field is a solution of the field
equations for the Lagrangian density $L_\mathrm{Dir}^+(\zeta)$ or the
field equations for the Lagrangian density $L_\mathrm{Dir}^-(\zeta)$.
\end{theorem}

\textbf{Proof\ }
Observe that the Dirac Lagrangian densities
$L_\mathrm{Dir}^\pm$ defined by formula~(\ref{Quasi-stationary case equation 2})
possess the property of scaling covariance:
\begin{equation}
\label{proof of theorem equation 1}
L_\mathrm{Dir}^\pm(e^h\zeta)=e^{2h}L_\mathrm{Dir}^\pm(\zeta)
\end{equation}
where $h:M\to\mathbb{R}$ is an arbitrary scalar function.
We claim that the statement of the theorem
follows from (\ref{lemma 2 equation 1})
and
(\ref{proof of theorem equation 1}).
The proof presented below is an abstract one and does not depend
on the physical nature of the
dynamical variable $\zeta$, the only requirement being that it is an element
of a vector space so that scaling makes sense.

Note that formulae
(\ref{lemma 2 equation 1})
and
(\ref{proof of theorem equation 1})
imply that the Lagrangian density $L$
possesses the property of scaling covariance,
so all three of our Lagrangian densities,
$L$, $L_\mathrm{Dir}^+$ and $L_\mathrm{Dir}^-$,
have this property.
Note also that if $\zeta$ is a solution of the field equation for
some Lagrangian density $\mathcal{L}\,$
possessing the property of scaling covariance
then $\mathcal{L}(\zeta)=0$. Indeed, let us perform a scaling
variation of our dynamical variable
\begin{equation}
\label{scaling variation}
\zeta\mapsto\zeta+h\zeta
\end{equation}
where $h:M\to\mathbb{R}$ is an arbitrary ``small'' scalar function
with compact support. Then
$0=\delta\int\mathcal{L}(\zeta)=2\int h\mathcal{L}(\zeta)$
which holds for arbitrary $h$ only if $\mathcal{L}(\zeta)=0$.

In the remainder of the proof the variations of $\zeta$ are arbitrary
and not necessarily of the scaling type (\ref{scaling variation}).

Suppose that $\zeta$ is a solution of the field equation for
the Lagrangian density $L_\mathrm{Dir}^+$.
[The case when $\zeta$ is a solution of the field equation for
the Lagrangian density $L_\mathrm{Dir}^-$ is handled similarly.]
Then $L_\mathrm{Dir}^+(\zeta)=0$ and, in view of (\ref{density is nonzero}),
$L_\mathrm{Dir}^-(\zeta)\ne0$.
Varying $\zeta$, we get
\begin{multline*}
\!\!\!\!
\delta\!\int\!\!L(\zeta)
=-\frac{32\omega}9\Bigl(
\int\!
\frac{L_\mathrm{Dir}^-(\zeta)}
{L_\mathrm{Dir}^+(\zeta)\!-\!L_\mathrm{Dir}^-(\zeta)}
\delta L_\mathrm{Dir}^+(\zeta)
+\!
\int\!\!
L_\mathrm{Dir}^+(\zeta)
\delta\frac{L_\mathrm{Dir}^-(\zeta)}
{L_\mathrm{Dir}^+(\zeta)\!-\!L_\mathrm{Dir}^-(\zeta)}
\Bigr)
\\
=\frac{32\omega}9\int\delta L_\mathrm{Dir}^+(\zeta)
=\frac{32\omega}9\,\delta\int L_\mathrm{Dir}^+(\zeta)
\end{multline*}
so
\begin{equation}
\label{formula for variation of our action}
\delta\int L(\zeta)=\frac{32\omega}9\,\delta\int L_\mathrm{Dir}^+(\zeta)\,.
\end{equation}
We assumed that $\zeta$ is a solution of the field equation for
the Lagrangian density $L_\mathrm{Dir}^+$ so
$\delta\int L_\mathrm{Dir}^+(\zeta)=0$ and formula
(\ref{formula for variation of our action}) implies that
$\delta\int L(\zeta)=0$. As the latter is true for an
arbitrary variation of $\zeta$ this means that
$\zeta$ is a solution of the field equation for the Lagrangian
density $L$.

Suppose that $\zeta$ is a solution of the field equation for
the Lagrangian density $L$.
Then $L(\zeta)=0$ and formula (\ref{lemma 2 equation 1})
implies that either $L_\mathrm{Dir}^+(\zeta)=0$ or $L_\mathrm{Dir}^-(\zeta)=0$;
note that in view of (\ref{density is nonzero}) we cannot have simultaneously
$L_\mathrm{Dir}^+(\zeta)=0$ and $L_\mathrm{Dir}^-(\zeta)=0$.
Assume for definiteness that $L_\mathrm{Dir}^+(\zeta)=0$.
[The case when $L_\mathrm{Dir}^-(\zeta)=0$ is handled similarly.]
Varying $\zeta$ and repeating the argument from the previous paragraph
we arrive at (\ref{formula for variation of our action}).
We assumed that $\zeta$ is a solution of the field equation for
the Lagrangian density $L$ so
$\delta\int L(\zeta)=0$ and formula
(\ref{formula for variation of our action}) implies that
$\delta\int L_\mathrm{Dir}^+(\zeta)=0$. As the latter is true for an
arbitrary variation of $\zeta$ this means that
$\zeta$ is a solution of the field equation for the Lagrangian
density~$L_\mathrm{Dir}^+$. $\square$

\

The proof of Theorem \ref{main theorem} presented above may appear
to be non-rigorous but it can be easily recast in terms of
explicitly written field equations.

\

Theorem \ref{main theorem} establishes that in the quasi-stationary case our
model reduces to a pair of massless Dirac equations (\ref{Dirac equation}).
There is, however, a small logical flaw in this statement.
The full time-dependent field equations for our model (as well as
for the massless Dirac model) are obtained by varying the spinor
field $\xi$ by a $\delta\xi$ with arbitrary dependence on time
whereas in the proof of Theorem \ref{main theorem} we have effectively varied the
spinor field $\xi$ maintaining quasi-stationarity, i.e. we took
\begin{equation}
\label{Quasi-stationary case equation 99}
\delta\xi^a(x^0,x^1,x^2,x^3)=
e^{-i\omega x^0}\delta\zeta^a(x^1,x^2,x^3)
\end{equation}
(compare with (\ref{Quasi-stationary case equation 1})). Note that
the $\delta\zeta$ in the above formula does not depend on time. If
we now modify (\ref{Quasi-stationary case equation 99}) so that
$\delta\zeta$ depends on time this will generate an extra term in
the field equations, one with the time derivative. It turns out that
this extra term with the time derivative vanishes. For the sake of
brevity we do not present the corresponding calculation.

\section{Plane wave solutions}
\label{Plane wave solutions}

Suppose that $M=\mathbb{R}^3$ is Euclidean 3-space equipped
with Cartesian coordinates $x=(x^1,x^2,x^3)$
and metric
$g_{\alpha\beta}=\operatorname{diag}(-1,-1,-1)$.
Let us choose constant Pauli matrices
\begin{equation}
\label{Plane wave solutions equation 1}
\sigma_{\alpha a\dot b}=
\begin{pmatrix}
\sigma_{1a\dot b}\\
\sigma_{2a\dot b}\\
\sigma_{3a\dot b}
\end{pmatrix}
:=
\begin{pmatrix}
\begin{pmatrix}
0&1\\
1&0
\end{pmatrix}
\\
\begin{pmatrix}
0&i\\
-i&0
\end{pmatrix}
\\
\begin{pmatrix}
1&0\\
0&-1\end{pmatrix}
\end{pmatrix}
\end{equation}
(compare with
(\ref{special formula for Pauli matrices}),
(\ref{Pauli matrices s}))
and seek solutions of the form
\begin{equation}
\label{Plane wave solutions equation 2}
\xi^a(x^0,x^1,x^2,x^3)=
e^{-i(\omega x^0+k\cdot x)}\zeta^a
\end{equation}
(compare with
(\ref{Quasi-stationary case equation 1}))
where $\omega\ne0$ is a real number,
$k=(k_1,k_2,k_3)$ is a real constant covector
and $\zeta\ne0$ is a (complex) constant spinor.
We shall call solutions of the type
(\ref{Plane wave solutions equation 2}) \emph{plane wave}.
In seeking plane wave solutions
what we are doing is separating out all the variables,
namely, the time variable $x^0$ and the spatial variables
$x=(x^1,x^2,x^3)$.

We look at the field equations of our model described in Section
\ref{Our model}. These field equations are highly nonlinear so it is
not \emph{a priori} clear that one can seek solutions in the form of plane
waves. However, plane wave solutions are a special case of
quasi-stationary solutions and the latter were analysed in Section
\ref{Quasi-stationary case}. We know
(Theorem \ref{main theorem}) that
in the quasi-stationary case our
model reduces to a pair of Dirac equations (\ref{Dirac equation}).
Substituting
(\ref{Notation and conventions equation 1}),
(\ref{Plane wave solutions equation 1})
and
(\ref{Plane wave solutions equation 2})
into (\ref{Dirac equation}) we get
\begin{equation}
\label{Plane wave solutions equation 3}
\begin{pmatrix}
\pm\omega-k_3&-k_1-ik_2\\
-k_1+ik_2&\pm\omega+k_3
\end{pmatrix}
\begin{pmatrix}
\zeta^1\\
\zeta^2
\end{pmatrix}=0.
\end{equation}
The determinant of the matrix in the LHS of
(\ref{Plane wave solutions equation 3}) is
$\omega^2-k_1^2-k_2^2-k_3^2$ so this system has a nontrivial solution
$\zeta$ if and only if $k_1^2+k_2^2+k_3^2=\omega^2$. Our model
is invariant under rotations of the Cartesian coordinate system
(orthogonal transformations of the coordinate system which preserve
orientation), so, without loss of generality,
we can assume that
\begin{equation}
\label{Plane wave solutions equation 4}
k_\alpha=
\begin{pmatrix}
0\\
0\\
\pm\omega
\end{pmatrix}.
\end{equation}
Substituting (\ref{Plane wave solutions equation 4}) into
(\ref{Plane wave solutions equation 3}) we conclude that,
up to scaling by a nonzero complex factor,
we have
\begin{equation}
\label{Plane wave solutions equation 5}
\zeta^a=
\begin{pmatrix}
1\\
0
\end{pmatrix}.
\end{equation}

Combining formulae (\ref{Plane wave solutions equation 2}),
(\ref{Plane wave solutions equation 4}) and
(\ref{Plane wave solutions equation 5})
we conclude that for each
real $\omega\ne0$ our model admits, up to a rotation
of the coordinate system and rescaling,
two plane wave solutions and that these plane wave solutions are given by the explicit
formula
\begin{equation}
\label{Plane wave solutions equation 6}
\xi^a=
\begin{pmatrix}
1\\
0
\end{pmatrix}e^{-i\omega(x^0\pm x^3)}.
\end{equation}

Let us now rewrite the plane wave solutions
(\ref{Plane wave solutions equation 6})
in terms of our original dynamical variables, coframe $\vartheta$
and density $\rho$. Substituting
(\ref{Notation and conventions equation 1}),
(\ref{Plane wave solutions equation 1})
and
(\ref{Plane wave solutions equation 6})
into
(\ref{common language equation 1})--(\ref{common language equation 5})
we get $\rho=1$ and
\begin{equation}
\label{Plane wave solutions equation 7}
\vartheta^1_\alpha
=\begin{pmatrix}
\cos2\omega(x^0\pm x^3)\\
-\sin2\omega(x^0\pm x^3)\\
0
\end{pmatrix},
\quad
\vartheta^2_\alpha
=\begin{pmatrix}
\sin2\omega(x^0\pm x^3)\\
\cos2\omega(x^0\pm x^3)\\
0
\end{pmatrix},
\quad
\vartheta^3_\alpha
=\begin{pmatrix}
0\\
0\\
1
\end{pmatrix}.
\end{equation}
Note that scaling of the spinor $\zeta$ by a
nonzero complex factor is equivalent to scaling
of the density $\rho$ by a positive real factor and
time shift $x^0\mapsto x^0+\operatorname{const}$.

We will now establish how many different (ones that cannot be
continuously transformed into one another) plane wave solutions we
have. To this end, we rewrite formula
(\ref{Plane wave solutions equation 7}) in the form
\begin{equation}
\label{Plane wave solutions equation 8}
\vartheta^1_\alpha
=\!\begin{pmatrix}
\cos2|\omega|(x^0+b x^3)\\
-a\sin2|\omega|(x^0+b x^3)\\
0
\end{pmatrix}\!,
\
\vartheta^2_\alpha
=\!\begin{pmatrix}
a\sin2|\omega|(x^0+b x^3)\\
\cos2|\omega|(x^0+b x^3)\\
0
\end{pmatrix}\!,
\
\vartheta^3_\alpha
=\!\begin{pmatrix}
0\\
0\\
1
\end{pmatrix}
\end{equation}
where $a$ and $b$ can, independently, take values $\pm1$.
It may seem that we have a total of 4 different plane wave solutions.
Recall, however (see Remark \ref{rigid rotations of coframe}), that we can
perform rigid rotations of the coframe and that we have agreed to view
coframes that differ by a rigid rotation as equivalent.
Let us perform a rotation of the coordinate system
\[
\begin{pmatrix}
x^1\\
x^2\\
x^3
\end{pmatrix}
\mapsto
\begin{pmatrix}
x^2\\
x^1\\
-x^3
\end{pmatrix}
\]
simultaneously with a rotation of the coframe
\[
\begin{pmatrix}
\vartheta^1\\
\vartheta^2\\
\vartheta^3
\end{pmatrix}
\mapsto
\begin{pmatrix}
\vartheta^2\\
\vartheta^1\\
-\vartheta^3
\end{pmatrix}.
\]
It is easy to see that the above transformations
turn a solution of the form
(\ref{Plane wave solutions equation 8})
into a solution of this form again only with
\[
a\mapsto-a,
\qquad
b\mapsto-b.
\]
Thus, the numbers $a$ and $b$ on their own do not characterise
different plane wave solutions. Different plane wave solutions are
characterised by the number $c:=ab$ which can take two values, $+1$
and $-1$.

The bottom line is that we have two essentially different types of
plane wave solutions. These can be written, for example, as
\begin{equation}
\label{Plane wave solutions equation 9}
\vartheta^1_\alpha
\!=\!\begin{pmatrix}
\cos2|\omega|(x^0+x^3)\\
\mp\sin2|\omega|(x^0+x^3)\\
0
\end{pmatrix}\!,
\quad
\vartheta^2_\alpha
\!=\!\begin{pmatrix}
\pm\sin2|\omega|(x^0+x^3)\\
\cos2|\omega|(x^0+x^3)\\
0
\end{pmatrix}\!,
\quad
\vartheta^3_\alpha
\!=\!\begin{pmatrix}
0\\
0\\
1
\end{pmatrix}\!.
\end{equation}

The plane wave solutions (\ref{Plane wave solutions equation 9})
describe travelling waves of rotations. Both waves travel with
the same velocity in the negative $x^3$-direction.
The difference between the two solutions is in the direction of rotation
of the coframe: if we fix $x^3$ and look at the evolution of
(\ref{Plane wave solutions equation 9}) as a function of time $x^0$ then
one solution describes a clockwise rotation whereas the other solution
describes an anticlockwise rotation.
We identify one of the solutions
(\ref{Plane wave solutions equation 9})
with a left-handed neutrino and
the other with a right-handed antineutrino.

\section{Discussion}
\label{Discussion}

As explained in Section \ref{Our model}, our mathematical model is a
special case of the theory of teleparallelism which in turn is a
special case of Cosserat elasticity.

The differences between our mathematical model
formulated in Section \ref{Our model} and
mathematical models commonly used in teleparallelism are as follows.
\begin{itemize}
\item
We assume the metric to be prescribed (fixed) whereas in
teleparallelism it is traditional to view the metric as a dynamical
variable. In other words, in works on teleparallelism it is
customary to view (\ref{constraint for coframe}) not as a constraint
but as a definition of the metric and, consequently, to vary the
coframe without any constraints at all. This is not surprising as
most, if not all, authors who contributed to teleparallelism came to
the subject from General Relativity.
\item
We choose a very particular Lagrangian density (\ref{Our model equation 6})
containing only one irreducible piece of torsion (axial) whereas in
teleparallelism it is traditional to choose a more general
Lagrangian containing all three pieces
(tensor, trace and axial), see formula (26)
in \cite{cartantorsionreview}.
In choosing our
particular Lagrangian density (\ref{Our model equation 6})
we were guided by the principle of conformal invariance.
\end{itemize}

The main result of our paper is Theorem \ref{main theorem}
which establishes that in the
quasi-stationary case (prescribed oscillation in time with frequency $\omega$)
our mathematical model is equivalent to a pair
of massless Dirac equations (\ref{Dirac equation}).

This leaves us with two issues.
\begin{itemize}
\item[A]
What can be said about the general
case, when the the spinor field $\xi$ is an arbitrary function
of all spacetime coordinates $(x^0,x^1,x^2,x^3)$
and is not necessarily of the form (\ref{Quasi-stationary case equation 1})?
\item[B]
What can be said about the relativistic version of our model
described in Section \ref{Relativistic}?
\end{itemize}
The two issues are, of course, related: both arise because in formulating
our basic model in Section \ref{Our model} we adopted the Newtonian approach
which specifies the time coordinate $x^0$ (``absolute time'').

We plan to tackle issue A by means of perturbation theory. Namely,
assuming the metric to be flat (as in Section \ref{Plane wave solutions}),
we start with a plane wave (\ref{Plane wave solutions equation 2})
and then seek the unknown spinor field $\xi$ in the form
\begin{equation}
\label{Discussion equation 1}
\xi^a(x^0,x^1,x^2,x^3)=
e^{-i(\omega x^0+k\cdot x)}\zeta^a(x^0,x^1,x^2,x^3)
\end{equation}
where $\zeta$ is a slowly varying spinor field. Here ``slowly
varying'' means that second derivatives of $\zeta$ can be
neglected compared to the first. Our conjecture is that the application
of a formal perturbation argument will yield a massless Dirac equation for
the spinor field $\xi$.

We plan to tackle issue B by means of perturbation theory as well.
The relativistic version of our model has 3 extra field equations
corresponding to the 3 extra dynamical degrees of freedom
(Lorentz boosts in 3 directions). Our conjecture is that if we take
a solution of the nonrelativistic problem which is a perturbation of a plane
wave (as in the previous paragraph) then, at a perturbative level, this solution
will automatically satisfy the 3 extra field equations. In other words, we conjecture
that our nonrelativistic model possesses
relativistic invariance at the perturbative level.

The detailed analysis of the two issues flagged up above will be the
subject of a separate paper.

\end{document}